\documentclass[10pt,conference,nocompress]{IEEEtran}
\pdfoutput=1

\IEEEoverridecommandlockouts
\usepackage[switch]{lineno}
\usepackage{hyperref}
\hypersetup{hidelinks}
\usepackage{graphicx} 
\usepackage{multirow}
\usepackage{tabularx}
\usepackage{url}
\usepackage{cite}
\usepackage{pifont}
\usepackage{xcolor}
\usepackage{enumerate}
\usepackage{hyperref}
\usepackage{enumitem}
\usepackage{caption}
\begin{document}
\author{
    \IEEEauthorblockN{
    Qi Guo\textsuperscript{1}\IEEEauthorrefmark{1},
    Xiaofei Xie\textsuperscript{2},
    Shangqing Liu\textsuperscript{3}\IEEEauthorrefmark{2},
    Ming Hu\textsuperscript{2},
    Xiaohong Li\textsuperscript{1}\IEEEauthorrefmark{2},
    and Lei Bu\textsuperscript{3}
    \thanks{\IEEEauthorrefmark{1} This work was done during Qi Guo's visit to Singapore Management University.}
    \thanks{\IEEEauthorrefmark{2} Shangqing Liu and Xiaohong Li are corresponding authors.}}
    \IEEEauthorblockA{
    \textsuperscript{1} Tianjin University, P.R. China\\
    \textsuperscript{2} Singapore Management University, Singapore\\
    \textsuperscript{3} State Key Laboratory for Novel Software Technology, Nanjing University, P.R. China\\
    }
    \IEEEauthorblockA{bxguoqi@tju.edu.cn,
    xfxie@smu.edu.sg,
    shangqingliu@nju.edu.cn,\\
    ecnu\_hm@163.com,
    xiaohongli@tju.edu.cn,
    bulei@nju.edu.cn
    }
}
\makeatletter
\patchcmd{\@maketitle}
  {\addvspace{0.5\baselineskip}\egroup}
  {\addvspace{-1\baselineskip}\egroup}
  {}
  {}
\makeatother

\title{Intention is All You Need: Refining Your Code from Your Intention}
\newcommand{\fei}[1]{{\textcolor{red}{Fei:#1}}}
\newcommand{\sql}[1]{\textcolor{red}{sql: #1}}
\newcommand{\guo}[1]{\textcolor{cyan}{#1}}
\newcommand{\major}[1]{\textcolor{black}{#1}}
\maketitle

\begin{abstract}


Code refinement aims to enhance existing code by addressing issues, refactoring, and optimizing to improve quality and meet specific requirements. As software projects scale in size and complexity, the traditional iterative exchange between reviewers and developers becomes increasingly burdensome. While recent deep learning techniques have been explored to accelerate this process, their performance remains limited, primarily due to challenges in accurately understanding reviewers' intents.

This paper proposes an \textit{intention}-based code refinement technique that enhances the conventional \textit{comment-to-code} process by explicitly extracting reviewer intentions from the comments.
Our approach consists of two key phases: Intention Extraction and Intention Guided Revision Generation. Intention Extraction categorizes comments using predefined templates, while Intention Guided Revision Generation employs large language models (LLMs) to generate revised code based on these defined intentions. Three categories with eight subcategories are designed for comment transformation, which is followed by a hybrid approach that combines rule-based and LLM-based classifiers for accurate classification. Extensive experiments with five LLMs (GPT4o, GPT3.5, DeepSeekV2, DeepSeek7B, CodeQwen7B) under different prompting settings demonstrate that our approach achieves 79\% accuracy in intention extraction and up to 66\% in code refinement generation. Our results highlight the potential of our approach in enhancing data quality and improving the efficiency of code refinement.
\end{abstract}
\begin{IEEEkeywords}
code refinement, intention-based generation, large language model
\end{IEEEkeywords}
\section{Introduction}

Code refinement focuses on enhancing existing code by addressing potential issues, refactoring, and optimizing to improve quality and meet specific requirements~\cite{bacchelli2013expectations,rigby2013convergent}. It is particularly crucial in software development as projects grow larger and more complex~\cite{fagan2002history,mcintosh2014impact}. 
According to an official report~\cite{Autio2022}, code review and refinement remain the most effective strategies for companies to improve code quality~\cite{sadowski2018modern,czerwonka2015code}. Thus, automated code refinement is crucial for software development~\cite{bavota2015four}.

Traditionally, the code refinement process involves two sequential phases~\cite{bosu2013impact,rigby2014peer,bosu2015characteristics}. Initially, code reviewers examine the code and provide feedback (i.e., review comments) on any identified issues. Subsequently, software developers modify the code based on these comments. This process is iterative, with reviewers continuously identifying issues in subsequent refinements until the code is considered issue-free. Reviewers and developers often rely on natural language for communication, which can lead to a semantic gap that complicates their interactions, making the process both cumbersome and time-consuming~\cite{yang2016mining}.

A significant amount of work~\cite{tufano2019learning, tufano2021towards, tufano2022using, li2022automating} has been conducted to expedite this process. Despite these efforts, the performance has remained limited. For instance, the state-of-the-art technique only has 40\% accuracy on the popular code refinement benchmark~\cite{li2022automating}. Recent works~\cite{guo2024exploring, pornprasit2024fine,pornprasit2024gpt,yang2023enhancing,almeida2024aicodereview,fan2024exploring,watanabe2024use} have been exploring the feasibility of using large language models (LLMs) for code refinement. However, these techniques merely employ LLMs in a straightforward manner without designing effective solutions tailored to the specific characteristics of this task. As a result, they also fail to achieve satisfactory performance.

The primary issue with current research lies in the proposed techniques failing to accurately comprehend the reviewer's \textit{intent}, which is critical for code refinement. The reasons can be primarily attributed to two aspects. Some reviewers write comments that are very vague and brief, such as suggesting to delete a line of code without specifying the exact line number. As a result, models fail to learn the reviewer's precise intent. Guo et al.~\cite{guo2024exploring} further indicate that when the comments lack explicit location information or clear modification strategies, even models like ChatGPT~\cite{ouyang2022training,achiam2023gpt} underperform due to their inability to understand the intention. On the other hand, the models currently used in code refinement have limited learning capabilities and cannot handle complex scenarios effectively. Recent work~\cite{tufano2024code} shows that current code-pretrained models, such as T5CR~\cite{tufano2022using} and CodeReviewer~\cite{li2022automating}, are effective in automating basic code formatting improvements but struggle with more complex logical code modifications. These models often overfit to simple refinements, such as renaming a variable from \textit{a} to \textit{b}, while misinterpreting intricate tasks as trivial ones.

To address the aforementioned issues, this paper proposes an \textit{intention}-based code refinement technique. Specifically, we extract the \textit{intention} from comments—a structured and templated summary of the reviewer’s modification intent. A clearer extracted intention (e.g., an instruction to change $a$ to $b$) leads to a more accurate code refinement. Unlike conventional end-to-end code refinement, which involves complex understanding challenges, our approach decomposes the process into two sequential phases: Intention Extraction and Intention-Guided Code Modification Generation. In the first phase, a predefined template is used to extract the intention from the reviewer’s comments. In the second phase, large language models (LLMs) and rule-based methods generate revised code based on the extracted intention, rather than relying solely on potentially vague reviewer comments. \major{
To define the intention templates for extracting intentions from comments, we conducted a preliminary study on 1,100 commits from GitHub. These commits were classified into three major categories based on their intentions: explicit change suggestions, reversion suggestions, and other general suggestions. While the first two categories convey clear intentions, we further summarized six intention templates to capture the broader scope of general suggestions. To accurately classify the comment category and extract the intention from a given review comment, we propose a hybrid approach that combines rule-based methods with LLM-based classifiers.
}
After the intentions are extracted, we apply different prompting strategies, including simple prompts, retrieval-augmented generation (RAG) prompts, and self-generated prompts, to explore the effect of various prompts for LLMs in the code refinement process.

To evaluate the effectiveness of our proposed approach, we selected \textbf{5} typical LLMs, including GPT4o~\cite{achiam2023gpt}, GPT3.5~\cite{ouyang2022training}, DeepSeekV2~\cite{zhu2024deepseek}, DeepSeek7B~\cite{guo2024deepseek}, and CodeQwen7B ~\cite{bai2023qwen} for the experiments. Extensive experiments confirmed that our approach can achieve an accuracy of \textbf{79}\% in intention extraction, with a maximum accuracy of \textbf{66}\% in generating revised code. Furthermore, different prompt strategies exhibit varying performance across different models in code refinement. In general, the performance of RAG prompts is stable. Lastly, larger models with more excellent reasoning capabilities demonstrated more significant improvements, with GPT4o and DeepSeekV2 showing accuracy enhancements of 8 and 9 percentage points, respectively. The main contributions of our paper are summarized as follows:

\begin{itemize}[leftmargin=*,topsep=2pt]
    \item We reformulate the comment-to-code refinement process into two sequential steps: \textit{comment-to-intent} and \textit{intent-to-code} generation. This decomposition reduces task complexity and improves LLMs' ability to understand reviewer intent.
    \item We propose a hybrid approach incorporating rule-based and LLM-based classifiers to extract the reviewer's \textit{intention} accurately.
    \item Extensive experiments demonstrate the effectiveness of our approach, achieving \textbf{79}\% accuracy in intention extraction and up to \textbf{66}\% accuracy in generating revised code.
\end{itemize}

\section{Background and Motivation}

\subsection{Code Refinement}\label{sec:background}

In the code review process, the developer first modifies the initial version of the code, resulting in a second version.  The modifications made to the code are captured as the last code diff hunk (denoted as \texttt{LastCodeDiffHunk}). The reviewer then evaluates the last code diff hunk along with the second version of the code to provide review comments (denoted as  \texttt{ReviewComment}). Each review comment is linked to a specific line in the code (denoted as \texttt{ReviewLine}). Subsequently, the developer modifies the second version of the code based on the reviewer’s comments, generating the third version of the code that meets the reviewer’s requirements. This subsequent modification process is referred to as the code refinement task. For clarity, the second version is henceforth referred to as the \texttt{OriginalCode}, and the third version as the \texttt{RevisedCode}.


The traditional code refinement, which we refer to as Basic Code Refinement, involves taking \texttt{OriginalCode} and \texttt{ReviewComment} as input and producing \texttt{RevisedCode} as output. However, we found that in some tasks, \texttt{ReviewLine} and \texttt{LastCodeDiffHunk} are needed to help understand the \texttt{ReviewComment}. 
We define two new tasks to address these issues: Position-Aware Code Refinement, which takes \texttt{OriginalCode}, \texttt{ReviewLine}, \texttt{ReviewComment} as input; and Comprehensive Code Refinement, which takes \texttt{OriginalCode}, \texttt{ReviewLine}, \texttt{ReviewComment} and \texttt{LastCodeDiffHunk} as input.

\subsection{Motivation}

Current approaches to code refinement are typically framed as end-to-end tasks, where machine learning models are expected to generate revised code directly from input data. This process often resembles a black-box guesswork for the models, lacking a complete reasoning process. 
We draw inspiration from the approaches employed by human developers to tackle code refinement tasks, enhancing the reasoning capabilities of large models by decomposing the complex task into subtasks: intention understanding and code refinement guided by the extracted intention.




\subsection{Data Collection}

Currently, there are two widely used datasets in the domain of code refinement: the Tufano dataset~\cite{tufano2019learning} and the CodeReview dataset~\cite{li2022automating}. As previously discussed, our research necessitates that the dataset provides the following five fields: \texttt{OriginalCode}, \texttt{ReviewLine}, \texttt{ReviewComment}, \texttt{LastCodeDiffHunk} and \texttt{RevisedCode}. The Tufano dataset lacks the \texttt{LastCodeDiffHunk} field and does not provide links to the original data. Although the CodeReview dataset lacks the \texttt{ReviewLine} and \texttt{LastCodeDiffHunk} fields, it does provide links to the original data. Consequently, we have chosen to use the CodeReview dataset and supplement the missing fields.

First, we describe the method for obtaining the \texttt{ReviewLine} field. We observed that by utilizing the GitHub REST API to retrieve code review information, it is possible to obtain a partial last code diff hunk (referred to as the diff hunk field in the API's JSON response\footnote{https://api.github.com/repos/meganz/sdk/pulls/comments/326107667}). We term this a partial last code diff hunk because it only provides the modification information preceding the review comment. 
Based on this pattern, we can deduce that the review line corresponds to the last line of the partial last code diff hunk.

Subsequently, we need to design a method to obtain the complete \texttt{LastCodeDiffHunk}. We observed that a single code review might encompass multiple commits reviewed by the reviewer. If a given pull request comprises \(n\) commits (\(commit_1, commit_2, ..., commit_n\)), the reviewer may choose to review all commits from \(commit_m\) to \(commit_n\) (where \(m \leq n\)), and then provide a review comment based on the accumulated file changes. 
We sequentially traverse all preceding commits in reverse order and compare them with \(commit_n\) to obtain the code diff hunk near the review line. By comparing this with the partial last code diff, we can identify the complete last code diff hunk that matches the partial last code diff hunk. Based on this method, we can derive the \texttt{LastCodeDiffHunk}.

\section{Intention-Based Framework}


\begin{figure*}[!t]
\centering
\includegraphics[width=\linewidth]{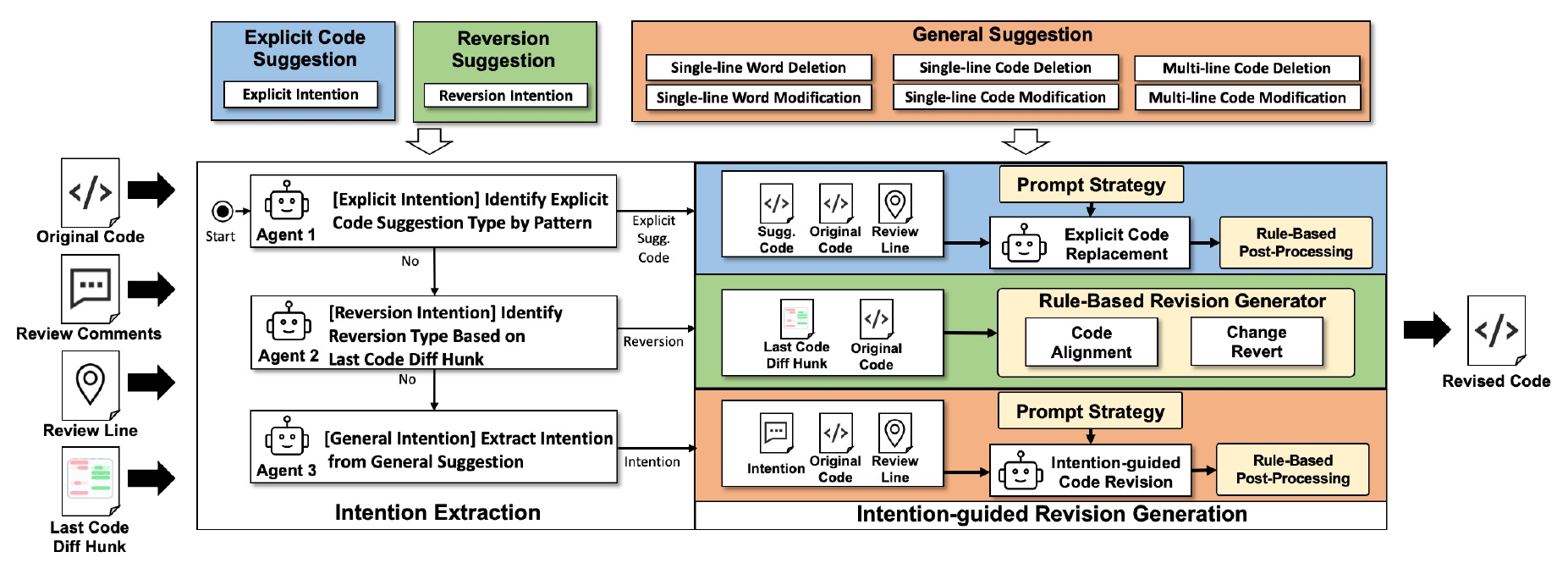}
\vspace{-6mm}
\caption{The framework of our intention-based code refinement.}
\label{fig:framework}
\vspace{-4mm}
\end{figure*}

\subsection{\major{Understanding Intention in Code Review}}
When developers tackle code refinement tasks, they begin by analyzing the reviewer's intention, particularly code modification suggestions. While the \texttt{ReviewComment} often provides these suggestions, it may also contain unrelated information such as explanations of reasons, polite expressions, or even emojis. Additionally, \texttt{ReviewComment} may include ambiguous references, colloquial language, or domain-specific terminology. Developers need to navigate these elements to extract the precise intention, which can often be summarized as a clear and templated directive, such as \texttt{"revert the previous modification"} or \texttt{"replace word A with word B"}. \major{Thus, understanding the review comment and distilling the intended action (i.e., intention) are crucial for performing the correct code refinement.}

\major{
\emph{Preliminary Study on Review Comments and Intentions.} 
To understand the intentions behind code review comments, we conducted a preliminary human study by manually analyzing these comments.
Specifically, we selected 30 highly starred projects on GitHub and extracted code reviews from the pull requests submitted within each project.
Projects without suitable pull requests or code reviews were excluded. To ensure diversity across projects, we limited the sampling to a maximum of 100 code reviews per project. Finally, we collected a total of 1,100 code review instances from 15 projects.}

\major{Three coauthors conducted the study to understand the intentions behind code review comments. Based on this understanding, we developed a classification framework to categorize these comments according to their intentions. Specifically, our classification is guided by three key criteria: 1) the representativeness of each category, ensuring that each type of comment constitutes a significant portion of the dataset; 2) an intention-oriented approach designed to capture the underlying intentions behind each type of comment; and 3) the feasibility of reliably extracting these intentions from only code reviews. In cases of disagreement during the categorization process, another coauthor facilitated discussions to resolve differences until a consensus was reached. Through this collaborative process, we identified three categories of code review comments, each reflecting a distinct intention: \textit{Explicit Code Suggestions}, \textit{Reversion Suggestions}, and \textit{General Suggestions}. These categories are representative and account for 13\%, 23\%, and 64\% of the total comments, respectively. The details of these three categories and their associated intentions are summarized as follows:
}

\begin{itemize}[leftmargin=*,topsep=2pt]
    \item \textbf{Explicit Code Suggestions}: This category includes the exact code that needs to be applied. Developers only need to identify the appropriate location and insert the given suggestion code accordingly. \major{\emph{For this type of comment, the intention is clear and direct, as the target changes are already explicitly stated within the comment.}}

    \item \textbf{Reversion Suggestions}: These imply that the reviewer explicitly or implicitly indicates that the previous modification was unsuitable and suggests reverting to the version before the last modification. \major{\emph{For this type of comment, the intention is to restore the code to its previous state.}}
    
    \item \textbf{General Suggestions}: \major{Comments that do not fall into the two specific categories above are classified as general suggestions, which lack explicit intentions. For these comments, we characterize their intention based on two general aspects: the type of change (i.e., insertion, modification, and deletion) and the corresponding scope of the change (i.e., single-line or multi-line). Since insertion is a specific form of modification, we group it under the broader category of changes, resulting in four general intention categories: single-line change, single-line deletion, multi-line change, and multi-line deletion. Additionally, we observed that word-level changes are particularly common in single-line edits. Therefore, we distinguish between word-level changes and code-level changes within single-line edits, which maintains clarity and operability. \emph{Ultimately, for this type of comment, we have six types of general intentions: }
    }
      \begin{itemize}
        \item Single-line change: Change word \texttt{(...)} to \texttt{(...)}
        \item Single-line change: Delete word \texttt{(...)}
        \item Single-line change: Change the code to \texttt{<code>}
        \item Single-line change: Delete code \texttt{<code>}
        \item Multi-line change: Delete code lines \texttt{<code>}
        \item Multi-line change: Change the code lines \texttt{<code 1>} to \texttt{<code 2>}
    \end{itemize}
\end{itemize}

\major{
Note that our classification differs from those in previous works~\cite{tufano2024code, fregnan2022happens} due to a difference in focus. Previous studies primarily aim to perform post-analysis, understanding the concrete actions from both refined code and comments, such as renaming variables or fixing specific bugs. In contrast, our work focuses on pre-analysis to identify the potential intention from review comments, which is more difficult. While it is possible to refine these classifications and intentions further, doing so would significantly increase the complexity of predicting such intentions from code review. An incorrect prediction of intention could potentially misguide the refinement process, leading to unintended changes or deviations from the desired outcomes. Therefore, this paper focuses on intentions that are either easy to extract (i.e., explicit changes and reversions) or with high-level patterns (i.e., the six general intention patterns). Our results (see Section~\ref{sec:rq2}) demonstrate the effectiveness of this approach. More fine-grained intention extraction and analysis are left as our future work.
}




\subsection{Intention Extraction}


Based on the intentions in different kinds of code reviews, we design the corresponding methods to extract intention and generate code revision. As illustrated in Fig.~\ref{fig:framework}, we employ three agents to distinguish among \textit{Explicit Code Suggestions}, \textit{Reversion Suggestions}, and \textit{General Suggestions}.

\textit{Intention for Explicit Code Suggestions.} In Agent 1, we utilize a rule-based method to identify the most explicit category: \textit{Explicit Code Suggestions}, which shows the direct and explict intention. 
This category includes cases where the review comment contains suggestion code snippets in the format \verb|```suggestion <code>```|. Accompanying these suggestions, there may be explanatory remarks and other exchanges between the reviewer and developer. We use regular expressions to determine if a case contains a suggestion code. If it does, it is categorized as \textit{Explicit Code Suggestions}.

\textit{Intention for Reversion Suggestions.} For cases not classified as \textit{Explicit Code Suggestions}, we employ Agent 2 to determine whether they fall under Reversion Suggestions. This assessment primarily relies on the \texttt{LastCodeDiffHunk}, which captures modifications from the Pre-Modification Code to the Post-Modification Code (also serving as the Original Code). We analyze whether the last modification involved an addition, deletion, or revision, and design specific prompts to evaluate whether the \texttt{ReviewComment} implies or explicitly requests a reversion of these changes.

If the \texttt{LastCodeDiffHunk} involves a deletion, we match the review comment into one of the following comment types:
The review comment with one of the following intentions:
 \begin{itemize}[leftmargin=*,topsep=2pt]
        \item  Expressing an opinion: You shouldn't delete this code.
        \item Expressing an opinion: You still need this code.
        \item Expressing an opinion: Change the code back.
        \item  Raising a question: Why delete this code?
        \item  Raising a question: Why did you do this?
        \item Giving a suggestion: You should add another piece of code.
\end{itemize}
Our preliminary study shows that the first five types cover the main deletion-related reversions, while the last one indicates no reversion. To accurately match with the types, we design a single-choice prompt for LLM-based classification, reducing ambiguities caused by variations in phrasing and tone. This approach improves the effectiveness of identifying and distinguishing reversion intents for deletions. Detailed prompts, along with handling for additions and revisions, are available on our website~\cite{IntentionWebsite}.

\begin{figure}[!t]
\centering
\includegraphics[width=.8\linewidth]{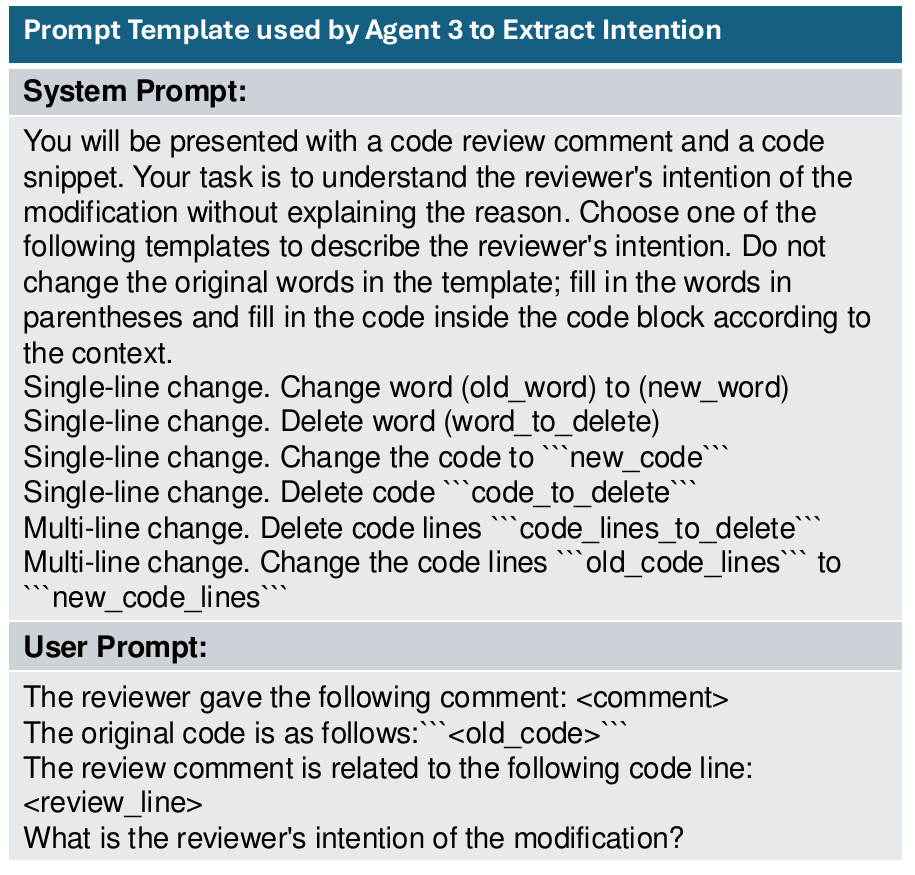}
\vspace{-2mm}
\caption{\major{The prompt used for extracting general intention.}}
\label{fig:agent3}
\vspace{-4mm}
\end{figure}

\major{
\textit{Intention for General Suggestions.} If the code reviews are not matched with the explicit suggestions and reversion suggestions, they will be categorized as \textit{General Suggestions}. We employ Agent 3 to further classify them into six subcategories: four types of single-line modifications and two types of multi-line modifications.
As shown in Fig.~\ref{fig:agent3}, the templates used by Agent 3 consist of two components: the \textit{System} and \textit{User} prompts. The \textit{System} prompt instructs the LLM to interpret the code review according to one of the six predefined Intention templates. The models used in our experiments support \textit{System} prompts, a parameter commonly available in most modern LLMs.
Notably, each intention category template includes placeholders that the LLM must fill based on the specific case details in the user prompt.
This design clarifies the reviewer's intent and facilitates the subsequent generation of revised code.}




\subsection{Intention Guided Revision Generation}

\begin{figure*}[!t]
\centering
\includegraphics[width=0.85\linewidth]{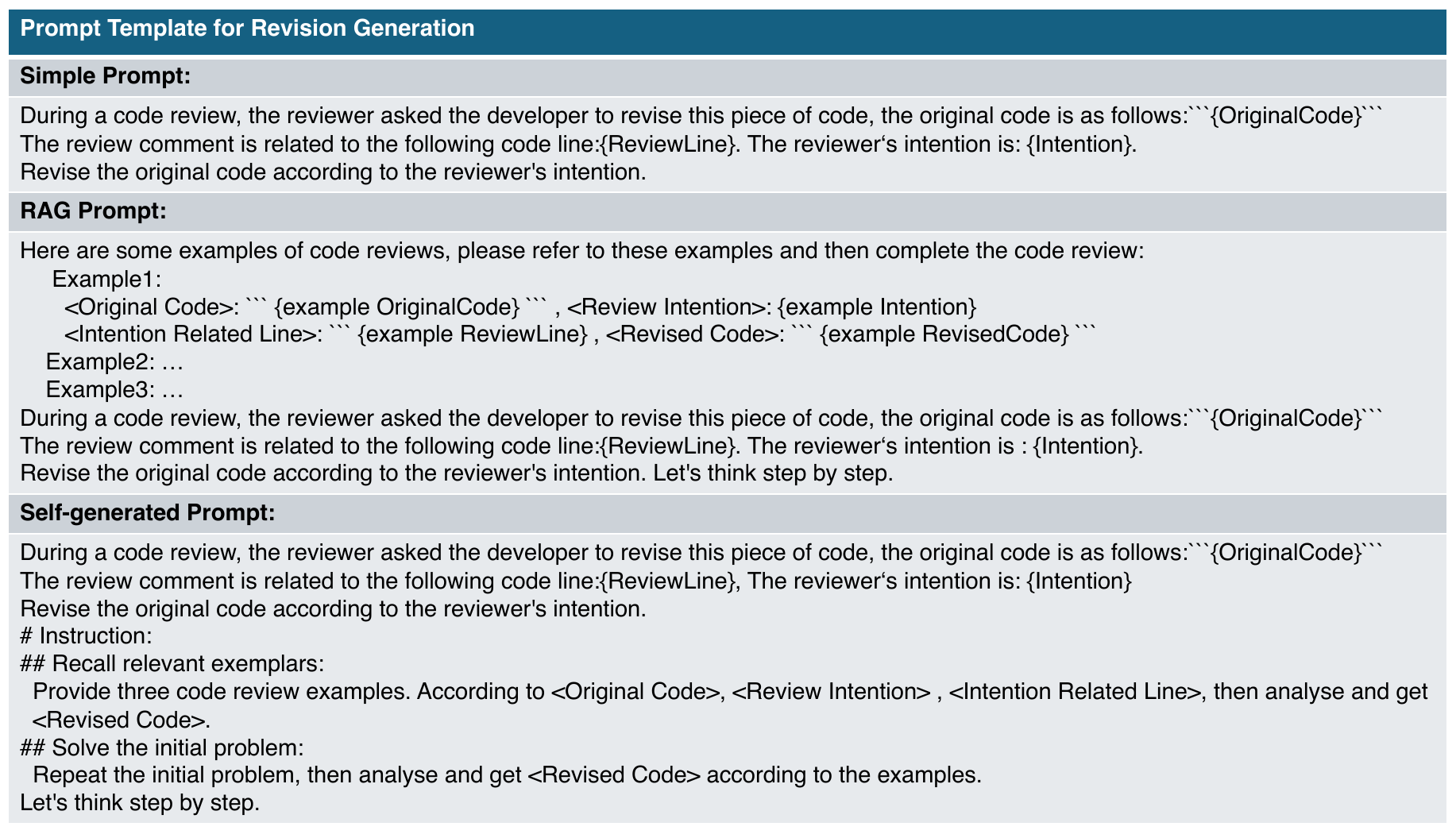}
\vspace{-2mm}
\caption{The format of the used prompt in the generation.}
\label{fig:framework_prompt}
\vspace{-4mm}
\end{figure*}

As illustrated in Fig.~\ref{fig:framework}, the Intention Guided Revision Generation process consists of two primary steps: \textit{generation} and \textit{post-processing}. Each of the three distinct intention categories—Explicit Code Suggestions, Reversion Suggestions, and General Suggestions—follows its own specific generation and post-processing. Post-processing is used to adjust and repair the generated code in cases where LLMs may over-modify or fail to maintain consistency.

For the Explicit Code Suggestions and General Suggestions categories, the generation phase is similar. Both utilize LLMs as the foundational method for code generation. Our framework does not restrict the prompting strategy for LLMs. Various prompt strategies could be incorporated within our framework. We have implemented three commonly used prompt strategies in code tasks: Simple Prompts, RAG (Retrieval-Augmented Generation) Prompts, and Self-generated Prompts. Below, we describe each of these prompt design strategies, as illustrated in Fig.~\ref{fig:framework_prompt}:

\noindent \textbf{Simple Prompt:} This strategy involves describing the task scenario and introducing each field's information, instructing the model to make modifications as required.

\noindent \textbf{RAG Prompt:} \major{This approach enhances few-shot prompting by selecting relevant examples from a retrieval database. The database consists of key-value pairs, where the retrieval key is a combination of the intention and the comment, and the value includes \texttt{OriginalCode}, \texttt{Intention}, \texttt{ReviewLine}, and \texttt{RevisedCode}. For a new case, the comment and its associated intention are used to retrieve relevant examples, which are then appended to the front of a simple prompt, creating a tailored few-shot prompt.}

\noindent \textbf{Self-Generated Prompt:} This approach allows the model to generate code refinement examples on its own. Each example includes the \texttt{OriginalCode}, \texttt{Intention}, \texttt{ReviewLine}, and \texttt{RevisedCode}. The model then uses these self-generated examples as inspiration to address the original problem.

Next, we describe the generation and post-processing for each Intention category in detail:

\noindent \textbf{Explicit Code Suggestions:} We input the \texttt{OriginalCode}, \texttt{SuggestionCode}, and \texttt{ReviewLine} in the model. \texttt{SuggestionCode} refers to the code suggested within the \texttt{ReviewComment}. Based on these inputs, we design specialized post-processing criteria tailored to the unique characteristics of code refinement:

\begin{itemize}[leftmargin=*,topsep=2pt]
    \item Inclusion of Suggestion Code: The suggestion code must appear in the revised code.
    \item Invariant Code Context: The original code’s preceding and succeeding segments should remain unchanged, with the suggestion code either inserted in the middle or replacing a middle segment.
\end{itemize}

Using these characteristics, we first locate the suggestion portion within the revised code. If there is a complete match of the suggestion code in the revised code, the location is successfully identified. If not, the suggestion code has not been fully replicated. By determining the maximum matching probability for each line, we can locate the corresponding suggestion code section and copy the complete suggestion code. Additionally, by applying the rule that each line of revised code should originate from either the original code or the suggestion code, we trim redundant sections and fill in missing parts of the revised code.

\noindent \textbf{Reversion Suggestions:} For Reversion Suggestions, we employ a rule-based approach to generate the revision, eliminating the need for additional post-processing steps. \major{Specifically, this involves reverting the changes that transformed the Pre-Modification Code into the Post-Modification Code (which also serves as the Original Code), resulting in the revised code.} We begin by aligning the \texttt{LastCodeDiffHunk} with the \texttt{OriginalCode} and then revert the previous code changes to the \texttt{OriginalCode}: delete added lines and add deleted lines from the previous modifications. This rule-based generation approach negates the need for any further post-processing.

\noindent \textbf{General Suggestions:} For General Suggestions, we first employ LLMs to generate the code based on \texttt{OriginalCode}, \texttt{Intention} and \texttt{ReviewLine}. We then design two specific rules for the post-processing of the refined code:

\begin{itemize}[leftmargin=*,topsep=2pt]
    \item Comment Consistency: If the intention suggests new code without comments, the modified code should also lack comments.
    \item  Line Consistency: For single-line modification, other lines should remain unchanged.
\end{itemize}

\section{Experimental Setup}
To evaluate the effectiveness of the proposed approach, we design the following research questions.
\begin{itemize}[leftmargin=*,topsep=2pt]
    \item \major{RQ1: How accurate are LLMs in extracting intentions?}
    \item RQ2: To what extent is the Intention-based framework effective in code refinement?
    \item RQ3: To what extent does the framework improve performance across different intention categories?
    \item RQ4: To what extent is intention-based dataset cleaning effective?
\end{itemize}





 

\subsection{RQ1: Intention Extraction Accuracy}
The accuracy of the intention understanding is important for code refinement. 
To measure the accuracy, we constructed a test dataset from the existing code review dataset~\cite{li2022automating}. We randomly select 2,000 samples for manual annotation to assess the accuracy of different LLMs in intention extraction. 

As shown in Fig.~\ref{fig:framework}, for reversion intention and general intention, we use the LLM to extract the intention. To check the accuracy of the extracted intention, we invited two PhD candidates to conduct manual annotations. In cases of differing results, the annotators reached a consensus through discussion to finalize the annotations. During the annotation process, we not only assessed the correctness of intention extraction but also filtered out invalid data. Specifically, cases where the revised code was unrelated to the review comments were excluded to ensure that only relevant and high-quality data were used in subsequent research questions.

\begin{figure}[!t]
\centering
\includegraphics[width=0.85\linewidth]{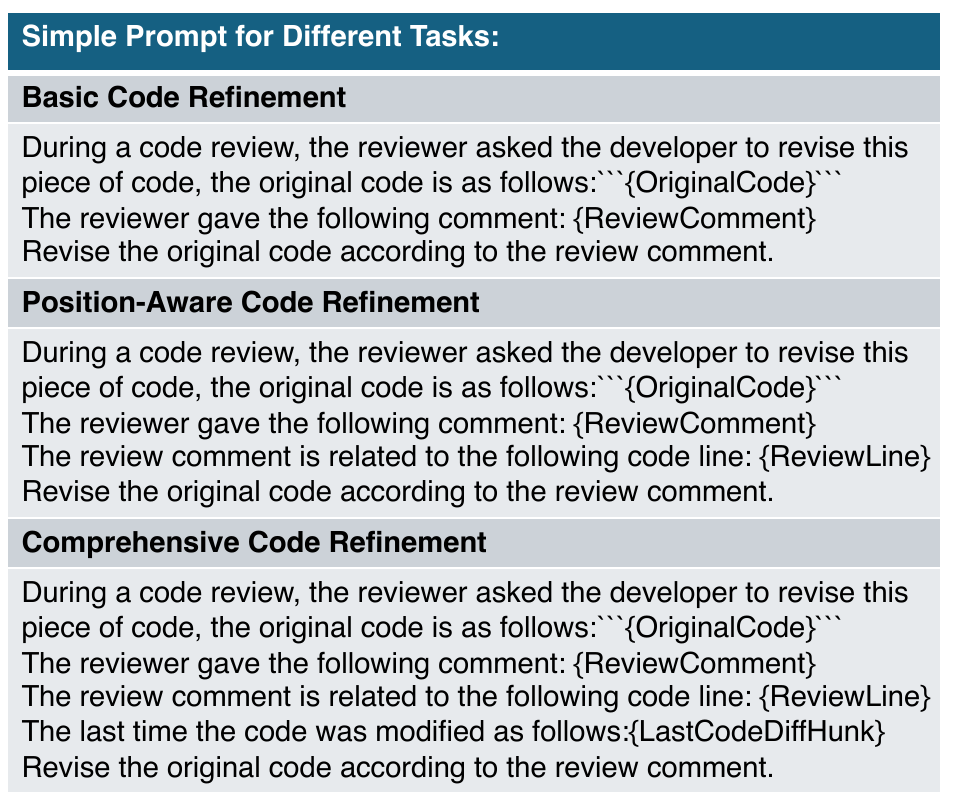}
\vspace{-2mm}
\caption{The used prompt format for different tasks.}
\label{fig:prompt2}
\vspace{-4mm}
\end{figure}

\subsection{RQ2: Intention-based Framework Effectiveness}
To evaluate the effectiveness of our proposed intention-based code refinement, we select two state-of-the-art baselines for comparison. 



\noindent \textbf{CodeReviewer~\cite{li2022automating}:} 
It designed three pre-training objectives, i.e., Diff Tag Prediction, Denoising Objective, and Review Comment Generation, to pre-train the model based on CodeT5~\cite{codet5} for code review activities. Several downstream tasks, including code change quality estimation, code review generation and code refinement, are selected to evaluate the effectiveness of the proposed models.

\noindent \textbf{T5CR~\cite{tufano2022using}:} It utilized two datasets including the official Stack Overflow dump (i.e., SOD) and CodeSearchNet (i.e., CSN) to pre-train the code review model based on T5 architecture. A tokenizer, i.e., SentencePiece~\cite{kudo2018sentencepiece}, is adopted to tokenize the source code, and the input sequence's maximum length is increased to 512 for training. 

Apart from these baselines, we also comprehensively evaluate the effectiveness of different LLMs and prompt strategies. In particular, we select three closed-source LLMs i.e., GPT-4o-2024-05-13 (GPT4o)~\cite{achiam2023gpt}, GPT-3.5-turbo-0125 (GPT3.5)~\cite{ouyang2022training}, DeepSeek-Coder-V2-0724 (DeepSeekV2)~\cite{zhu2024deepseek} and two open-source large models: CodeQwen1.5-7B-Chat (CodeQwen7B)~\cite{bai2023qwen} and Deepseek-coder-6.7b-instruct (DeepSeek7B)~\cite{guo2024deepseek} for evaluation.
\major{Note that, since different models have varying capabilities in extracting intentions, we present results using both the intentions extracted by the model itself and those extracted by a high-quality model (i.e., GPT-4o). The use of GPT-4o intentions allows us to evaluate whether providing the correct intention can enhance performance in code refinement tasks.}




For LLMs, we selected different prompting strategies:

\noindent \textbf{Simple Prompt:} As shown in Fig.~\ref{fig:prompt2}, the Simple Prompt first describes the scenario, then introduces the input information in the task, and finally requests the generation of revised code based on the provided information. This is a concise and effective prompt design used in Guo et al.~\cite{guo2024exploring}.

\noindent \textbf{Simple COT Prompt:} This prompt builds upon the simple prompt by adding the phrase ``Let's think step by step." This technique is employed in model reasoning~\cite{wei2022chain,wang2022self}.

\noindent \textbf{Tufano COT Prompt:} It is introduced in Tufano et al.~\cite{tufano2024code}, which first requires determining which of the following six categories the modification belongs to before completing the modification and then utilizes LLM for the generation.








\noindent \textbf{Random Few-shot Prompt:} For each case, three data examples are randomly selected as examples for the prompt as few-shot (excluding self-examples). The data fields provided as examples depend on the task's input fields, such as \texttt{OriginalCode}, \texttt{ReviewComment}, \texttt{ReviewLine}, \texttt{RevisedCode} in Comprehensive Code Refinement, and only \texttt{OriginalCode}, \texttt{ReviewComment}, \texttt{RevisedCode} in Basic Code Refinement. After providing examples as prompt hints, they are concatenated with the simple prompt.


\noindent \major{\textbf{RAG Prompt:} 
To compare with the Random Few-Shot Prompt, we design a Retrieval-Augmented Generation (RAG) prompting method as an alternative approach to few-shot prompting, leveraging relevant example selection. The retrieval database is constructed from the dataset used in RQ1, specifically the 2,000 randomly selected samples from the CodeReview dataset. However, only 1,337 of these samples are included in the database, as the remaining 663 cases exhibit low-quality refinements that do not align well with the review comments.
We selected these samples for two reasons: 1) The CodeReview dataset uniquely provides the complete data required for this study, including \texttt{OriginalCode}, \texttt{ReviewComment}, \texttt{ReviewLine}, \texttt{RevisedCode}, and \texttt{LastCodeDiffHunk}. 2) The quality of the retrieval dataset is crucial, and these 2,000 samples have been manually analyzed in RQ1, ensuring their reliability.
During testing, for each test case, we retrieve three samples to construct a 3-shot prompt. If the test data is included in the retrieval results, it is excluded and replaced with another sample. To enhance contextual relevance, we use BM25 to select semantically similar examples, ensuring that the retrieved examples closely align with the test data's context..}

\noindent \textbf{Self-generated Prompt:} This prompt technique addresses mathematical and reasoning problems by utilizing the model's own understanding ability~\cite{yasunaga2023large}. It first generates several examples and then uses these examples to inspire the model to answer the original question.


\noindent \textbf{Evaluation Metrics:} To fully automate the code refinement task, we prioritize the exact match (EM) between the model's output and the actual revised code. While metrics like BLEU and Code-BLEU can reflect proximity to the ground truth, they do not effectively gauge the degree to which the model's output aids programmers in code refinement. Particularly in simple tasks, if the model's output is not completely correct, it may be less beneficial for programmers to use the model's output for refinement than to directly use the review comment. Therefore, our evaluation metric only includes the EM value.

\subsection{RQ3: Improvement Across Different Intention Categories}


In this experiment, we mainly evaluated the effects of different prompting strategies and three prompting strategies: Simple Prompt, RAG, and Self-generated Prompt. We aim to understand how much our framework improves performance for each of Intention types compared to methods that do not use the framework.

\major{Additionally, we conducted ablation experiments on the Intention Extraction component of our framework, specifically evaluating the impact of the three agents (Agent1, Agent2, and Agent3). For each agent, we tested the Exact Match (EM) values by removing the respective agent and comparing the results. This analysis allows us to quantify the individual contribution of each agent to the overall performance of the framework and to identify which components are most critical for improving the accuracy of intention-based predictions.}



\subsection{RQ4: Intention-Based Dataset Cleaning}

Lastly, we aim to explore the capability of using intention to enhance data quality. A significant challenge in code refinement data quality is the inconsistency between the content expressed in the review comment and the modifications made in the revised code. Currently, no practical method exists for aligning the review comment and revised code at the semantic level. We will attempt to use intention to determine whether the modifications in the revised code meet the reviewer's requirements.

The Intention Method we designed is as follows: we provide the model with information about the Intention, Original Code, Review Line, and Revised Code and use GPT4o as a classification model to determine whether the modifications in the Revised Code meet the Intention requirements.

We also compared this with a common method that does not use the intention: we provide the model with information about the Review Comment, Original Code, Review Line, and Revised Code and use GPT4o as a classification model to determine whether the revised code meets the requirements of the review comment.



\section{experimental results}

\subsection{RQ1: Intention Extraction Accuracy}

\begin{table}[!t]
\caption{\major{Results of intention accuracy.}}
\footnotesize
\centering
\label{tb:rq1-1}
\resizebox{.4\textwidth}{!}{
\begin{tabular}{ccccc}
\hline
           & Explicit & Reversion        & General          & All clean data   \\ \hline
\#Samples  & 175      & 308              & 854              & 1337             \\
\major{GPT3.5}     & -        & \major{84.42\%}          & \major{49.18\%}          & \major{63.95\%}          \\
GPT4o      & -        & \textbf{99.03\%} & \textbf{66.86\%} & \textbf{78.61\%} \\
\major{DeepSeekV2} & -        & \major{95.45\%}          & \major{62.88\%}          & \major{75.24\%}          \\
\major{DeepSeek7B} & -        & \major{80.19\%}          & \major{46.72\%}          & \major{61.41\%}          \\
\major{CodeQwen7B} & -        & \major{82.14\%}          & \major{43.91\%}          & \major{60.06\%}          \\ \hline
\end{tabular}
}
\vspace{-4mm}
\end{table}

\major{
We manually annotated the data to select valid samples, with 1,337 out of a total of 2,000 identified as valid. To ensure annotation quality, we enlisted two senior PhD candidates specializing in code learning to conduct the manual labeling. Disagreements between the annotators were observed in 249 instances, and inter-rater reliability was assessed using Cohen’s Kappa coefficient, which yielded a value of 0.719, indicating acceptable agreement.
Subsequently, we used various models to generate intentions using our framework, and then manually assessed each generated intention for its correctness. Notably, since the Explicit category is classified using predefined rules, we just measure the accuracy of the reversion intention and general intention. 
}

\major{As shown in Table~\ref{tb:rq1-1}, we observe that GPT-4o and DeepSeekV2 are significantly more effective than the other three models. Among these, Reversion Intention is relatively easier to extract, with all models achieving at least 80\% accuracy, while GPT-4o achieves an impressive 99.03\%.}

It is noteworthy that 20\% of Reversion Suggestions overlap with General Suggestions, representing a dual classification challenge. In these instances, predicting the intention as either category is considered correct. 
For General Suggestions cases, the highest accuracy is 66.86\%, indicating that this category presents more significant challenges for the model to comprehend. Unlike the other categories, General Suggestions lack specific patterns or structured cues, making it more difficult for the model to pinpoint the exact intention behind the suggestions. This lower accuracy highlights the complexity and ambiguity inherent in general suggestions, which often require a deeper understanding of the context and the underlying reasoning behind the recommendations.

\vspace{5pt}\noindent \fbox{
\parbox{0.95\linewidth}{\textbf{Answers to RQ1}: \major{GPT-4o achieves the highest accuracy in intention extraction, with DeepSeekV2 performing competitively (with 3\% less). General intentions, however, are more challenging to extract due to the inherent complexity and the often ambiguous nature in the review comments.}}
}

\begin{table*}[!t]
\caption{\major{Comparative results between our intention-based framework and LLM-based baselines.}}
\footnotesize
\label{tb:rq2-1}
\resizebox{.99\textwidth}{!}{
\begin{tabular}{ccccccccccccccccccccccccc|cc}
\hline
           & \multicolumn{3}{c}{Simple Prompt} &  & \multicolumn{3}{c}{Simple COT} &  & \multicolumn{3}{c}{Tufuno COT} &  & \multicolumn{3}{c}{Random fewshot} &  & \multicolumn{3}{c}{RAG} &  & \multicolumn{3}{c}{Self-generated} &  & \multicolumn{2}{c}{Intention   Framework with RAG} \\ \cline{2-4} \cline{6-8} \cline{10-12} \cline{14-16} \cline{18-20} \cline{22-24} \cline{26-27} 
           & Basic     & P-A       & Comp.     &  & Basic    & P-A      & Comp.    &  & Basic    & P-A      & Comp.    &  & Basic      & P-A       & Comp.     &  & Basic  & P-A    & Comp. &  & Basic      & P-A       & Comp.     &  & Same             & \major{GPT4o Inte.}            \\ \hline
GPT3.5     & 38.00     & 38.29     & 36.42     &  & 34.85    & 36.87    & 32.76    &  & 19.97    & 17.20    & 11.74    &  & 45.25      & 48.62     & 32.39     &  & 46.45  & 49.89  & 33.96 &  & 31.86      & 25.88     & 27.97     &  & 53.40                & \major{\textbf{58.86}}              \\
GPT4o      & 41.14     & 46.60     & 29.54     &  & 39.72    & 43.68    & 26.25    &  & 20.12    & 24.98    & 13.76    &  & 49.21      & 52.88     & 35.00     &  & 49.59  & 54.08  & 36.95 &  & 51.01      & 56.84     & 34.41     &  & 64.77                & \major{\textbf{64.77}}              \\
DeepSeekV2 & 34.63     & 38.07     & 27.60     &  & 41.29    & 46.37    & 31.49    &  & 24.31    & 27.60    & 17.43    &  & 46.45      & 52.66     & 42.41     &  & 48.09  & 55.35  & 45.62 &  & 44.95      & 53.03     & 38.37     &  & 64.25                & \major{\textbf{68.06}}              \\
DeepSeek7B & 28.50     & 34.48     & 24.38     &  & 29.92    & 36.28    & 24.01    &  & 5.76     & 6.24     & 3.60     &  & 33.28      & 39.49     & 25.58     &  & 36.05  & 41.59  & 25.88 &  & 7.26       & 8.68      & 5.01      &  & 49.29                & \major{\textbf{56.40}}              \\
CodeQwen7B & 26.10     & 33.13     & 19.22     &  & 25.06    & 31.04    & 16.98    &  & 2.85     & 3.29     & 1.97     &  & 35.75      & 40.99     & 17.80     &  & 35.83  & 44.05  & 20.12 &  & 13.69      & 19.67     & 7.93      &  & 48.47                & \major{\textbf{54.75}}              \\ \hline
\end{tabular}
}
\vspace{-4mm}
\end{table*}

\subsection{RQ2: Intention-based Framework Effectiveness}\label{sec:rq2}

\major{Table~\ref{tb:rq2-1} presents a comparison between our method and the LLM-based baselines. For the LLM baselines, we made extensive efforts to optimize their performance by setting multiple configurations, including six different prompts and three types of inputs (i.e., basic inputs, position-aware code refinement, and comprehensive code refinement; details are provided in Section~\ref{sec:background}).
For our method, we show the results of the RAG-based approach and the results with other prompts are shown in Table~\ref{tb:rq3-1}. Note that we present two types of results for our method: (1) using the same model for both intention extraction and code refinement (Column \textit{Same}) and (2) extracting accurate intentions with GPT-4o and generating revised code using other LLMs based on these intentions (Column \textit{GPT4o Inte.}).}

From Table~\ref{tb:rq2-1}, we can observe that using the Intention-based method improved the performance of all models compared to all baselines. The model with the highest improvement was DeepSeekV2, which saw an increase from 55.35\% to 64.25\%, a 9 percentage points improvement. The model with the least improvement was GPT3.5, which increased from 49.89\% to 53.40\%, a mere 3 percentage points improvement.  We provide several representative examples to demonstrate the advantages of the Intention-based framework, which are available on our website~\cite{IntentionWebsite}.

\noindent
\major{
\textbf{From the Perspective of Intention Quality}:
Comparing the results of our method using intentions generated by the model itself and those generated by GPT4o, it is evident that intention quality plays a critical role. While using the intentions generated by the same model (Column \textit{Same}) generally improves performance compared to LLM baselines, leveraging higher-quality intentions (Column \textit{GPT4o Inte.}) further enhances the results. This is due to the fact that some models, as shown in Table~\ref{tb:rq1-1}, are not as effective as GPT4o in extracting accurate intentions. Surprisingly, we also observed that DeepSeekV2 outperformed GPT4o (68.06 VS 64.77) when provided with the same intentions (from GPT4o). This demonstrates that, while DeepSeekV2 may not excel in intention extraction, it has superior code refinement capabilities when given accurate intentions. The results underscore the strength of our framework, which allows for separate optimization of the agents responsible for intention understanding and code refinement under a given intent. A similar trend is observed for other prompts, as shown in Table~\ref{tb:rq3-1}.
}

\noindent
\textbf{From the Perspective of Input Complexity:} By comparing the baseline results, we found that across all six prompt strategies, the Position-Aware Code Refinement task performed the best, followed by the Basic Code Refinement task, and lastly the Comprehensive Code Refinement task. Although the Comprehensive Code Refinement task provided the most information, including all necessary data, all models and prompt methods struggled to effectively utilize this information. This may be due to the complexity of understanding \texttt{LastCodeDiffHunk} and the potential for overly long prompts to reduce model focus, impairing comprehension of key information. Incomplete input information reduces model effectiveness, while excessive input information hampers task understanding and reduces effectiveness. The Position-Aware Code Refinement task, which includes \texttt{OriginalCode}, \texttt{ReviewComment}, and \texttt{ReviewLine}, balances the completeness of information with the model's processing capability (5 percentage points better than Basic Code Refinement and about 10 percentage points better than Comprehensive Code Refinement in general). 


\noindent
\textbf{From the Perspective of Prompt Strategy:} 
Except for the DeepSeekV2 model, other models showed only slight improvements with the Simple COT Prompt method over the Simple Prompt, and even showed declines for GPT3.5 and GPT4o in the Basic task. This indicates that for most models like GPT3.5, GPT4o, DeepSeek7B, and CodeQwen7B, Simple COT Prompt without clear step-by-step guidance does not significantly enhance code refinement tasks. However, for the DeepSeek model, using COT significantly improved accuracy in Basic, Position-Aware, and Comprehensive tasks by up to 8 percentage points, demonstrating DeepSeekV2's strong reasoning capabilities. 

Consistent with our intuition, the Random Few-Shot Prompt outperformed the Simple COT Prompt and Simple Prompt. Furthermore, RAG demonstrated superior performance compared to the Random Few-Shot Prompt across almost all models and tasks. \major{This indicates that retrieving more similar examples is highly beneficial, as it provides refinement guidance that aligns closely with both the format and the content of the task.
Therefore, we recommend that users intending to use the RAG prompt for code review tasks construct the retrieval database using a relevant historical code review dataset. Relevance can be assessed from various perspectives, including project similarity, task similarity, author alignment, dataset quality, and fine-grained intention matching. Specifically, retrieved data from the same or similar project, addressing similar tasks, or authored by the same individuals or the same group is likely to offer better guidance for new code refinement tasks, as these examples share meaningful similarities and contextual relevance.}

For Self-generation Prompt, compared to the RAG method, it showed a slight improvement for the GPT4o model, a slight decline for the DeepSeekV2 model, and a significant decline for the other three models. Notably, GPT4o and DeepSeekV2, the two largest models with the best overall performance in other prompt methods, benefited from Self-generation. This suggests that Self-generation Prompt is more effective for models with large parameter sizes and strong reasoning capabilities.

Last, we compared our method with pre-trained models, specifically CodeReviewer~\cite{li2022automating} and T5CR~\cite{tufano2022using}, which achieved 52.14\% and 18.84\% accuracy, respectively. As expected, our intention-based method demonstrates superior effectiveness, achieving significantly higher performance due to the use of LLMs and the incorporation of intention extraction (e.g., 64.77\% with GPT4o).



\vspace{5pt}\noindent \fbox{
\parbox{0.95\linewidth}{\textbf{Answers to RQ2}: The results show that the Intention-based framework outperforms the existing baselines, including both pre-trained models and LLMs with diverse prompts. 
\major{High-quality intentions play a critical role in achieving effective code refinement.}
}
}

\begin{table}[!t]
\centering
\caption{\major{Results of our method with different prompts.}}
\footnotesize
\label{tb:rq3-1}
\resizebox{.49\textwidth}{!}{
\begin{tabular}{cccclccc}
\hline
           & \multicolumn{3}{c}{Our Method (Same)}                    &  & \multicolumn{3}{c}{\major{Our Method (GPT4o   Inte.)}} \\ \cline{2-4} \cline{6-8} 
           & Simple Prompt  & RAG            & Self-generated &  & \major{Simple Prompt}  & \major{RAG}             & \major{Self-generated} \\ \hline
GPT3.5     & 53.03          & \textbf{53.40} & 43.38          &  & \major{57.14}          & \major{\textbf{58.86}}  & \major{46.60}          \\
GPT4o      & 64.10          & 64.77          & \textbf{65.97} &  & \major{64.10}          & \major{64.77}           & \major{\textbf{65.97}} \\
DeepSeekV2 & 60.88          & \textbf{64.25} & 61.03          &  & \major{61.03}          & \major{\textbf{68.06}}  & \major{64.10}          \\
DeepSeek7B & \textbf{50.04} & 49.29          & 26.55          &  & \major{54.45}          & \major{\textbf{56.40}}  & \major{27.75}          \\
CodeQwen7B & 46.00          & \textbf{48.47} & 29.32          &  & \major{50.11}          & \major{\textbf{54.75}}  & \major{27.45}          \\ \hline
\end{tabular}
}
\vspace{-4mm}
\end{table}

\begin{table*}[!t]
\caption{\major{Ablation results on different intention-based components.}}
\footnotesize
\label{tb:rq3-3}
\resizebox{.99\textwidth}{!}{
\major{
\begin{tabular}{cccccccccccccccc}
\hline
           & \multicolumn{3}{c}{Our Method (with Intention)}         &  & \multicolumn{3}{c}{w/o Explicit  Intention} &  & \multicolumn{3}{c}{w/o Reversion   Intention} &  & \multicolumn{3}{c}{w/o General Intention} \\ \cline{2-4} \cline{6-8} \cline{10-12} \cline{14-16} 
           & Simple Prompt & RAG            & Self-generated &  & Simple Prompt    & RAG             & Self-generated   &  & Simple Prompt    & RAG             & Self-generated    &  & Simple Prompt         & RAG                 & Self-generated        \\ \hline
GPT3.5     & 57.14         & \textbf{58.86} & 46.60          &  & 56.25(-0.9)      & 56.39(-2.47)    & 42.63(-3.96)     &  & 49.59(-7.55)     & 51.6(-7.26)     & 35.53(-11.07)     &  & 51.53(-5.61)          & 58.41(-0.45)        & 44.05(-2.54)          \\
GPT4o      & 64.10         & 64.77          & \textbf{65.97} &  & 64.03(-0.07)     & 63.65(-1.12)    & 65.45(-0.52)     &  & 59.61(-4.49)     & 60.21(-4.56)    & 61.41(-4.56)      &  & 57.37(-6.73)          & 62.3(-2.47)         & 65.75(-0.22)          \\
DeepSeekV2 & 61.03         & \textbf{68.06} & 64.10          &  & 60.58(-0.45)     & 65.52(-2.54)    & 61.41(-2.69)     &  & 55.27(-5.76)     & 61.85(-6.21)    & 56.32(-7.78)      &  & 52.35(-8.68)          & 63.95(-4.11)        & 63.05(-1.05)          \\
DeepSeek7B & 54.45         & \textbf{56.40} & 27.75          &  & 53.18(-1.27)     & 53.48(-2.92)    & 22.59(-5.16)     &  & 45.85(-8.6)      & 44.2(-12.19)    & 8.97(-18.77)      &  & 48.62(-5.83)          & 52.51(-3.89)        & 29.39(+1.65)          \\
CodeQwen7B & 50.11         & \textbf{54.75} & 27.45          &  & 46.97(-3.14)     & 52.81(-1.94)    & 19.75(-7.7)      &  & 38.0(-12.12)     & 46.6(-8.15)     & 6.96(-20.49)      &  & 47.8(-2.32)           & 54.75(0.00)          & 39.72(+12.27)         \\ \hline
\end{tabular}
}
}
\end{table*}

\subsection{RQ3: Ablation Study on Different Intention Categories}

\major{Table~\ref{tb:rq3-3} presents the ablation study results. The first column shows the performance with all intention-handling components included, while the subsequent columns illustrate the effects of removing intention-handling for each category (i.e., removing each agent in Fig.~\ref{fig:framework}). For example, removing the Explicit Intention component means that code reviews with explicit suggestions are instead handled by Agent 2 and Agent 3. Similarly, removing the Reversion Intention redirects code reviews with reversion suggestions to Agent 1 and Agent 3. For Agent 3, the general intention extraction is removed, and LLMs are used to handle these cases directly.
The results indicate that removing any of these intention components leads to performance drops across almost all scenarios, demonstrating the importance of each component. Notably, removing the Reversion Intention results in significant declines in EM scores across all models and prompt methods, with reductions ranging from 5 to 15 percentage points, highlighting its critical role in improving this category of code reviews.
}

\major{An exception is observed with the Self-generated prompt under the condition of ``w/o General Intention'', where performance increases slightly. This is likely because Self-generated prompts tend to perform worse with LLMs that have weaker reasoning capabilities, as discussed in RQ2. However, the Simple Prompt shows a significant drop, indicating that the intention-based method provides substantial improvements when using very basic prompts. In contrast, RAG-based and Self-generated prompts exhibit relatively smaller performance declines. These advanced prompts compensate for the models' limitations in implicitly understanding intentions, even when intentions are not explicitly provided.}

\begin{table*}[!t]
\caption{\major{Improvement results with different intention-based components.}}
\footnotesize
\label{tb:rq3-2}
\resizebox{.99\textwidth}{!}{
\begin{tabular}{cccccccccccc}
\hline
           & \multicolumn{3}{c}{Simple Prompt}           &  & \multicolumn{3}{c}{RAG}                     &  & \multicolumn{3}{c}{Self-generated}          \\ \cline{2-4} \cline{6-8} \cline{10-12} 
           & Explicit & Reversion & General (\major{Same/GPT4o}) &  & Explicit & Reversion & General (\major{Same/GPT4o}) &  & Explicit & Reversion & General (\major{Same/GPT4o}) \\ \hline
GPT3.5     & 29.71    & 62.01     & 2.86/\major{12.78}           &  & 8.57     & 57.47     & -9.57/\major{2.64}           &  & 38.29    & 70.13     & -1.29/\major{5.73}           \\
GPT4o      & 6.86     & 74.35     & 13.22/\major{13.22}          &  & 2.29     & 68.83     & 4.85/\major{4.85}            &  & 4.00     & 71.10      & 0.44/\major{0.44}            \\
DeepSeekV2 & 21.14    & 76.95     & 16.74/\major{16.45}          &  & 7.43     & 67.53     & 0.59/\major{7.49}            &  & 13.14    & 67.53     & -3.96/\major{1.47}           \\
DeepSeek7B & 25.71    & 65.91     & 2.71/\major{11.75}           &  & 14.29    & 60.71     & -6.14/\major{8.52}           &  & 38.86    & 73.38     & -5.43/\major{-3.23}          \\
CodeQwen7B & 29.71    & 67.21     & -3.43/\major{6.75}           &  & 24.00    & 57.14     & -12.00/\major{0.00}          &  & 50.29    & 68.51     & -19.86/\major{-22.91}        \\ \hline
\end{tabular}
}
\vspace{-4mm}
\end{table*}




\major{To further understand the improvements that are caused by our three intention-based components, we calculate the EM improvements on the test samples that are handled by each component, i.e., the EM difference compared to the best performance of the LLM baselines (see Table~\ref{tb:rq2-1}). Note that, due to that the intention quality is important, similar to RQ2, we show both results using intentions extracted by the same model and GPT4o. 
}


As shown in Table~\ref{tb:rq3-2}, the Intention-based framework demonstrates improvements in most cases. Notably, the most significant improvement is observed in the Reversion Suggestion category, with accuracy increases ranging from 57 to 76 percentage
points. This substantial gain is attributed to the rule-based refinement strategy employed for Reversion Suggestions, which ensures correct refinement when the given sample is accurately classified into this category during the intention analysis.

For the Explicit Code Suggestion category, our framework also shows improvement. For relatively weaker models like GPT3.5, DeepSeek7B, and CodeQwen7B, the improvement is particularly notable, often exceeding 20 percentage points, with a maximum increase of 50 percentage points. For stronger models like GPT4o and DeepSeekV2, it still provides slight improvements. 
\major{This is because explicit code suggestions are relatively easier for stronger models to handle, as the explicit intentions are more straightforward for them to extract. However, weaker models struggle to effectively extract these intentions and refine code, resulting in more significant improvements when using our method.}
\major{For tasks in the General Suggestion category, we observed that our method improves results when high-quality intention extraction (e.g., using GPT4o) is employed. However, when using the same model for both intention extraction and code refinement, the results decrease in some cases. This is because weaker models may extract incorrect intentions, which not only fail to assist the refinement process but can also lead to the generation of incorrect code.}

\major{Additionally, with the Simple Prompt strategy, the overall improvement is notable since the original model performs poorly, leaving substantial room for improvement. On the other hand, for RAG-based and Self-generated prompts, the overall performance is already enhanced, so the use of low-quality intentions (extracted by the same model) can more easily degrade performance.}

\major{
Furthermore, we found that the improvement in the General Suggestion category (even with high-quality intentions) is lower compared to Explicit Suggestion and Reversion Suggestion categories. This is because intentions in General Suggestions remain relatively high-level and less concrete, making it more challenging for the model to execute precise refinements compared to the other two categories.}


\vspace{5pt}\noindent \fbox{
\parbox{0.95\linewidth}{\textbf{Answers to RQ3}: \major{The results demonstrate that all three components contribute to performance improvement. However, their effectiveness varies. The Reversion Intention component achieves the largest improvement (over 50 percentage points) due to its rule-based code refinement, which ensures high accuracy. In contrast, the General Intention component shows relatively less improvement, as its intentions are less concrete compared to Explicit and Reversion Intentions, introducing larger uncertainty into the refinement process. Moreover, performance can decrease if the General Intention is not accurately extracted.}
}
}

\subsection{RQ4: Intention-Based Dataset Cleaning}

\begin{table}[!t]
\centering
\caption{Effectiveness on data cleansing.}
\scriptsize
\label{tb:rq4}
\begin{tabular}{ccccccc}
\hline
          & TP   & FP  & TN  & FN  & Accuracy & Precision \\ \hline
Intention-based & 1076 & 110 & 553 & 261 & 81.45\%  & 90.73\%   \\
Comment-based    & 1201 & 312 & 351 & 136 & 77.60\%  & 79.38\%   \\ \hline
\end{tabular}
\vspace{-4mm}
\end{table}

We further investigated the role of intention in data cleaning. As shown in Table~\ref{tb:rq4}, compared to directly using the review comment to verify whether the code modification meets the reviewer’s requirements, using Intention to verify the code modification proved to be more effective. The accuracy increased from 77.60\% to 81.45\%, and the precision increased from 79.38\% to 90.73\%.


Such improvements in accuracy and precision underscore the significance of utilizing intention as a guiding principle for code verification. Since dataset construction places a higher emphasis on data quality, the 12\% improvement in precision is significant for enhancing data quality. This increase ensures that the cleansed data is not only more accurate but also more reliable for downstream applications and analysis.

Overall, the intention-based approach demonstrates a more balanced and effective methodology for ensuring that code modifications align closely with the original reviewer's intentions, resulting in cleaner and more precise datasets. This shift toward a more intention-driven process marks a substantial advancement in the field of data cleansing, providing developers and data scientists with a more robust tool for maintaining code integrity and quality.

\vspace{5pt}\noindent \fbox{
\parbox{0.95\linewidth}{\textbf{Answers to RQ4}: The results indicate that intention-based cleansing is more effective for code refinement data cleansing than comment-based methods, achieving an accuracy of 81\% and a precision of 91\%.
}
}

\section{DISCUSSION}
\subsection{Discussions on Intention Classification and Extraction}

\major{Our framework divides the code refinement task into two steps: \textbf{intention analysis} and \textbf{intention-guided refinement}. This separation allows for improvements in effectiveness by enhancing both steps individually. Ideally, we aim to extract intentions as concretely as possible, such as Explicit and Reversion Intentions, which simplify the following refinement process. However, when the intention is less concrete (e.g., for General Suggestions), the refinement process involves more understanding difficulties, making performance improvements less significant.}

\major{
Our results also indicate that inaccurate intention extraction can degrade performance compared to an end-to-end refinement approach (see RQ3). This highlights why we only extract high-level intentions for General Suggestions, as it is challenging to ensure fully accurate and concrete intention extraction in these cases. Extracting incorrect intentions can lead to misguided refinements, which we aim to avoid.
}

\major{
In the future, refining the categories of General Intentions could further enhance the refinement process. For example, if the categories are more concrete, a rule-based method can be easily designed, or weaker models may better understand and refine the code. An ideal scenario would be that we have a complete classification of intention categories, where each category is sufficiently concrete to allow the use of reliable rule-based methods or even very weak models for effective refinement.
Explicit Suggestions and Reversion Suggestions are prime examples of such concrete categories.}

\subsection{Threats to Validity}
\noindent \textit{Model Threats:} We only tested the 7B versions of open-source code models, DeepSeek7B and CodeQwen7B, due to the resource limit. However, based on the performance of general open-source models like GPT4o and DeepSeekV2, our Intention-based framework performs excellently in code refinement tasks compared to other prompt techniques.
\major{Another potential threat is the length of the prompt templates. While longer prompts can pose input challenges for some models, the longest prompt template used in our study (the RAG prompt) contains only 141 tokens. The fields for each case (e.g., \texttt{OriginalCode}, \texttt{ReviewLine}, \texttt{Intention}) typically remain below 200 tokens. Even when including three-shot examples, the total input length remains under 1,000 tokens. Therefore, the prompt length is unlikely to affect the validity of our results.}

 \noindent \textit{Data Threats:} We only selected CodeReviewer dataset because other datasets lack some data fields and do not provide the link to the original data, making it impossible to use the Intention framework. However, the CodeReviewer dataset has been used in many papers and is recognized as a relatively complete and objective dataset. In the future, we will also try to collect more comprehensive and higher-quality datasets.
\major{We also acknowledge the potential risk of data leakage, particularly when using LLMs. While it is challenging to entirely rule out the possibility of data leakage within LLMs, our experimental results demonstrate that utilizing the Intention framework consistently yields better outcomes compared to LLMs not using it. This indicates that even in scenarios where data leakage may occur, the framework's design and methodology provide a significant performance advantage.}

 \noindent \textit{Efficiency Threats:} Our framework involves multiple LLM calls for classification and code generation, potentially making it slightly less efficient than other prompt strategies. However, the model’s classification response speed is relatively fast, and for Explicit Code Suggestions and Reversion Suggestions, we only need one step LLM call. Therefore, the overall impact on efficiency is not significant.



\section{related work}
Code refinement is a critical core component in the code review process, and numerous scholars have conducted research on the automation of code refinement. Tufano M.~\cite{tufano2019learning} were the pioneers in proposing the use of Neural Machine Translation (NMT) to learn the automated modification of Java methods based on review comments. Subsequently, Tufano R.~\cite{tufano2021towards} and Thongtanunam~\cite{thongtanunam2022autotransform} employed transformer~\cite{vaswani2017attention} models to train and enhance the original task’s performance. 
Tufano R.~\cite{tufano2022using} and Li~\cite{li2022automating} further advanced this field by using the Text-To-Text Transfer Transformer (T5) model~\cite{raffel2020exploring} and CodeT5 model~\cite{wang2021codet5}, pre-training code review-related tasks to enable the model to comprehend the meaning of the code and review comments. These approaches yielded significant improvements in downstream code refinement tasks.

With the rise of LLMs, many researchers have attempted to leverage them in software engineering \cite{ma2024specgen, ma2024speceval, kong2024contrastrepair, guo2024ft2ra, xia2023keep}. In particular, Guo~\cite{guo2024exploring} explored using ChatGPT for code refinement tasks, uncovering some prompt design techniques.
Tufano R.~\cite{tufano2024code} manually analyzed over 2,000 code refinement examples, evaluating three code review models~\cite{hong2022commentfinder, li2022automating, tufano2022using} and comparing their performance to ChatGPT. This analysis revealed that ChatGPT is highly competitive compared to previous methods. 
Pornprasit~\cite{pornprasit2024fine} experimented with various prompt strategies for LLMs in code refinement. They also fine-tuned ChatGPT using an API~\cite{ChatGPTblog}, enhancing the effectiveness of LLMs in this task.

Some studies have also involved classifying code refinement tasks.
Tufano~\cite{tufano2024code} categorizes code refinement based on the type of task and examines the performance of pre-trained models on different task types. 
Kononenko~\cite{kononenko2016code} studied the time and effort required by programmers for different types of code review tasks. Bacchelli~\cite{bacchelli2013expectations} investigated the categories of code review tasks, focusing on developer motivation and response speed. Pascarella~\cite{pascarella2018information} explored the information needed for different types of code refinement tasks, but their classification method is more oriented toward human understanding rather than guiding model modifications.



\section{Conclusion}
In this work, we propose an intention-based code refinement framework to enhance the effectiveness and accuracy of code refinement tasks. Our framework decomposes the process into two sequential phases: Intention Extraction and Intention-Guided Revision Generation. We evaluate the framework using five models, i.e., GPT-4o, GPT-3.5, DeepSeekV2, DeepSeek7B, and CodeQwen7B, across various prompting strategies. Extensive experimental results demonstrate that our approach outperforms both traditional methods and end-to-end LLM-based approaches. Additionally, we show that larger models with stronger reasoning capabilities achieve more significant improvements.

\section*{Acknowledgment}
This work was partially supported by the National Natural Science Foundation of China (Key Program, Grant No. 62332005), the National Research Foundation, Singapore, and the Cyber Security Agency under its National Cybersecurity R\&D Programme (NCRP25-P04-TAICeN). Lei Bu is supported in part by the Leading-edge Technology Program of Jiangsu Natural Science Foundation (No. BK20202001), the National Natural Science Foundation of China (No. 62232008, 62172200). Any opinions, findings and conclusions or recommendations expressed in this material are those of the author(s) and do not reflect the views of National Research Foundation, Singapore and Cyber Security Agency of Singapore.



\bibliographystyle{IEEEtran}
\bibliography{software}

\end{document}